\documentclass[prb,aps,twocolumn,superscriptaddress,showpacs,preprintnumbers,amsmath,amssymb]{revtex4}
\usepackage{color}
\usepackage{graphicx}
\usepackage{ulem}
\usepackage{hyperref}

\def\textbf#1{\boldsymbol{#1}}

\begin{document}

\title{Lattice dynamical signature of charge density wave formation in underdoped YBa$_{2}$Cu$_{3}$O$_{6+x}$}
\author{M.~Bakr}
\affiliation{Max-Planck-Institut~f\"{u}r~Festk\"{o}rperforschung,
Heisenbergstr.~1, D-70569 Stuttgart, Germany}
\affiliation{Physics Department, Jazan University, P.O. Box 114,
45142 Jazan, Kingdom of Saudi Arabia.}

\author{S. M.~Souliou}
\affiliation{Max-Planck-Institut~f\"{u}r~Festk\"{o}rperforschung,
Heisenbergstr.~1, D-70569 Stuttgart, Germany}

\author{S.~Blanco-Canosa}
\affiliation{Max-Planck-Institut~f\"{u}r~Festk\"{o}rperforschung,
Heisenbergstr.~1, D-70569 Stuttgart, Germany}
\affiliation{Helmholtz-Zentrum Berlin fr Materialien und Energie,
Albert-Einstein-Strasse 15, D-12489 Berlin, Germany}

\author{I.~Zegkinoglou}
\affiliation{Max-Planck-Institut~f\"{u}r~Festk\"{o}rperforschung,
Heisenbergstr.~1, D-70569 Stuttgart, Germany}
\affiliation{Chemical Sciences Division, Lawrence Berkeley National Laboratory, Berkeley, CA 94720, USA}

\author{H. Gretarsson}
\affiliation{Max-Planck-Institut~f\"{u}r~Festk\"{o}rperforschung,
Heisenbergstr.~1, D-70569 Stuttgart, Germany}

\author{J.~Strempfer}
\affiliation{Hamburger Synchrotronstrahlungslabor (HASYLAB)\\
 at Deutsches Elektronensynchrotron (DESY), 22605 Hamburg, Germany}

\author{T.~Loew}
\affiliation{Max-Planck-Institut~f\"{u}r~Festk\"{o}rperforschung,
Heisenbergstr.~1, D-70569 Stuttgart, Germany}

\author{C.T.~Lin}
\affiliation{Max-Planck-Institut~f\"{u}r~Festk\"{o}rperforschung,
Heisenbergstr.~1, D-70569 Stuttgart, Germany}

\author{R. Liang}
\affiliation{Department of Physics and Astronomy, University of British Columbia, Vancouver,V6T 1Z1, Canada}

\author{D. A. Bonn}
\affiliation{Department of Physics and Astronomy, University of British Columbia, Vancouver,V6T 1Z1, Canada}

\author{W. N. Hardy}
\affiliation{Department of Physics and Astronomy, University of British Columbia, Vancouver,V6T 1Z1, Canada}


\author{B.~Keimer}
\affiliation{Max-Planck-Institut~f\"{u}r~Festk\"{o}rperforschung,
Heisenbergstr.~1, D-70569 Stuttgart, Germany}

\author{M.~Le Tacon}
\affiliation{Max-Planck-Institut~f\"{u}r~Festk\"{o}rperforschung,
Heisenbergstr.~1, D-70569 Stuttgart, Germany}
\date{\today}

\begin{abstract}
We report a detailed Raman scattering study of the lattice dynamics in detwinned single crystals of the underdoped high temperature superconductor YBa$_{2}$Cu$_{3}$O$_{6+x}$ (x=0.75, 0.6, 0.55 and 0.45). Whereas at room temperature the phonon spectra of these compounds are similar to that of optimally doped YBa$_{2}$Cu$_{3}$O$_{6.99}$, additional Raman-active modes appear upon cooling below $\sim 170-200$ K in underdoped crystals. The temperature dependence of these new features indicates that they are associated with the incommensurate charge density wave state recently discovered using synchrotron x-ray scattering techniques on the same single crystals. Raman scattering has thus the potential to explore the evolution of this state under extreme conditions.
\end{abstract}


\maketitle

\section{Introduction}

When charges are injected into the Mott-insulating, antiferromagnetically ordered CuO$_2$ planes of layered copper oxides, antiferromagnetic long-range order is rapidly suppressed and is eventually replaced by high temperature superconductivity~\cite{Sachdev_RMP2003,Kivelson_RMP2003, Votja_AdvPhys2009}. At low doping levels, diffraction methods have revealed spatially periodic modulations of the charge density in the CuO$_2$ planes.
Two such patterns have been identified: the ``striped'' state in which antiferromagnetically ordered regions are separated by charged domain walls, which has been extensively studied in the `214' family [La$_{2-x}$Ba$_x$CuO$_4$ and La$_{2-x-y}$Sr$_x$(Nd,Eu)$_{y}$CuO$_4$] ~\cite{Fujita_JPSJ2012,Tranquada_Nature1995,Fujita_PRB2004,Abbamonte_NatPhys2005,Fink_PRB2009,Fink_PRB2011,Wu_NatureCommunications2012,
Wilkins_PRB2011,Dean_PRB2013}; and the biaxial charge density wave (CDW) state recently discovered in `123' compounds of composition (Y,Nd)Ba$_2$Cu$_3$O$_{6+x}$ \cite{Ghiringhelli_Science2012,Achkar_PRL2012,Blanco_PRL2013,Chang_NaturePhysics2012,Blackburn_PRL2013,Wu_Nature2011,Wu_Natcom2013, Thampy_PRB2013} and in Bi-based superconductors \cite{Comin,daSilva}. Whereas the data show clear evidence of competition between static stripe and CDW order and superconductivity, the influence of correlated charge fluctuations on $d$-wave Cooper pairing in the cuprates remains one of the central research questions in this field.

The spectroscopic determination of the dispersions and linewidths of lattice vibrations is a promising avenue to explore this issue, because they are expected to couple to charge fluctuations via the electron-phonon interaction~\cite{Reznik_PhysicaC2012}. Inelastic neutron scattering (INS) and inelastic x-ray scattering (IXS) studies have indeed revealed anomalous dispersions of the Cu-O-Cu bond-stretching and bond-bending vibrations of doped cuprates~\cite{Baron_JPCS2008,Uchiyama_PRL2004,Graf_PRB2007,Graf_PRL2008,Reznik_Nature2006,Pintschovius_PRB2004,Pintschovius_PRL2002,Raichle_PRL2011}. In the 214 system, the wave vector of the bond-stretching phonon anomalies detected by INS is compatible with stripe order. However, in order to provide a firm basis for the quantitative interpretation of these effects, it is desirable to identify clear manifestations of static stripe order in the electronic band dispersions and/or the phonon dispersions. In particular, one generally expects a Fermi surface reconstruction as well as new phonon modes at the Brillouin zone (BZ) center due to folding of the crystallographic BZ in the stripe-ordered state. However, to the best of our knowledge, neither of these effects has been clearly identified in stripe-ordered 214 materials.

Recent work on the 123 system has provided new perspectives on electronic ordering phenomena in the copper oxides. At low doping levels ($0.3 < x < 0.45$ in (Y,Nd)Ba$_2$Cu$_3$O$_{6+x}$)~\cite{Haug_NJP2010}, a unidirectional, static, incommensurate magnetic structure is observed, whereas at higher doping levels ($0.5 < x < 0.75$), a large gap opens in the spin excitations, and low energy fluctuations of the charge density appear~\cite{Ghiringhelli_Science2012}. The latter are found to be biaxial and incommensurate, and centered at a wave vector not directly related to the one of the magnetic fluctuations. Substitution of non-magnetic Zn for planar Cu revealed that static spin and charge modulated phases are actually competing in the 123 system~\cite{Blanco_PRL2013}.
The CDW correlation length increases strongly upon cooling and reaches $\sim 20$ lattice spacings at $T=T_c$, but the superconducting transition preempts the transition to CDW long-range order. Magnetic fields degrade superconductivity and enhance the CDW correlations, and recent evidence points towards the formation of a long-range ordered CDW state in high magnetic fields~\cite{Chang_NaturePhysics2012,Wu_Nature2011,Letacon_NatPhys2013}. The small Fermi surface pockets revealed by recent quantum oscillation measurements ~\cite{Doiron_Nature2007,Sebastian_RPP2012} have been attributed to electronic band-folding effects associated with CDW order. Very recent high-resolution IXS experiments have demonstrated static (albeit short-range) CDW order even in zero magnetic fields, presumably as a consequence of pinning of CDW fluctuations by defects~\cite{Blackburn_PRB2013,Letacon_NatPhys2013}.

In analogy to the 214 system, anomalies in the Cu-O-Cu bond-stretching and bond-bending vibrations have been observed at wave vectors compatible with those of the CDW fluctuations in 123 compounds. Pronounced anomalies were observed in acoustic and low-energy optical phonons at the CDW wavevector~\cite{Blackburn_PRB2013,Letacon_NatPhys2013}, thus providing a direct link between both phenomena. The situation regarding BZ-center phonons remains less clear, however, because numerous modes related to the CuO chains are present in both Raman and infrared spectra of underdoped 123 materials. In particular, a set of phonon modes detected by infrared spectroscopy at low temperatures had been tentatively attributed to charge correlations on the CuO chains~\cite{Bernhard_2002}. Since the newly discovered CDW can be unambiguously assigned to the CuO$_2$ planes \cite{Ghiringhelli_Science2012,Achkar_PRL2012,Blanco_PRL2013} and appears to be generic to the superconducting cuprates, a reinvestigation of this situation is in order.

In this paper, we present a systematic Raman scattering study of the lattice dynamics in the 123 family as function of doping, temperature and incident photon wavelength, using Raman scattering. This technique has proven to be a powerful probe of electronic and lattice vibrational excitations in strongly correlated systems~\cite{Devereaux_RMP2007}. In particular, the resulting data on 214 compounds at low frequencies have been interpreted in terms of phasons and amplitudons of charge stripes~\cite{Sugai_PRL2006}. In the same family, the symmetry-dependent Raman intensity has been attributed to collective electronic excitations arising from fluctuations of a charge-ordered state~\cite{Muschler_EPJ2010,Tassini_PRL2005,Caprara_PRL2005,Caprara_PRB2011}. Their absence in the 123 family has been discussed as a consequence of short correlation lengths~\cite{Tassini_PRB2008}.
Here, we checked five twin-free YBa$_{2}$Cu$_{3}$O$_{6+x}$ single crystals, with doping levels $p$ ranging from the optimal to the strongly underdoped, which have previously been investigated using resonant x-ray scattering~\cite{Ghiringhelli_Science2012, Achkar_PRL2012, Blanco_PRL2013}.
We show that despite the complexity of the Raman spectra of the underdoped, non-stoichiometric samples, which contain numerous features associated with chain oxygen ordering~\cite{Andersen_PhysicaC99,Zimmermann_PRB2003} and/or local defects, vibrational modes associated with the CDW can be unambiguously identified under specific resonant conditions. Raman scattering can thus be used as a sensitive and versatile tool to investigate CDW correlations in these compounds. In agreement with a previous study~\cite{Ghiringhelli_Science2012}, we observe that they disappear rapidly below $p \sim 0.09$ where the incommensurate spin fluctuations become static and where the quantum oscillations disappear. We further found that dynamical signatures of the CDW can be detected up to temperatures significantly higher than those at which signatures of static CDW order can no longer be discerned by x-rays. However, they disappear at temperatures below the pseudogap temperature $T^*$ as seen with various experimental techniques~\cite{Timusk_RPP99}.


\section{Experimental details}
\subsection{Samples}
The Raman experiments were performed on high quality detwinned YBa$_{2}$Cu$_{3}$O$_{6+x}$ single crystals grown as described in Refs.~\onlinecite{CTLin2002, Liang_PRB2006}. We present here measurements performed on five single crystals of compositions YBa$_{2}$Cu$_{3}$O$_{6.99}$ ($p \sim$ 0.16 and $T_c$=90~K), YBa$_{2}$Cu$_{3}$O$_{6.75}$ ($p \sim$ 0.135 and $T_c$=75~K), YBa$_{2}$Cu$_{3}$O$_{6.6}$ ($p \sim$ 0.12 and $T_c$=61~K), YBa$_{2}$Cu$_{3}$O$_{6.55}$ ($p \sim$ 0.10 and $T_c$=61~K) and YBa$_{2}$Cu$_{3}$O$_{6.45}$ ($p \sim$ 0.08 and $T_c$=35~K). The crystals were cut into uniform rectangular shapes of typical size 3$\times$3$\times$1 mm$^{3}$, and detwinned individually under a uniaxial pressure as described in Refs.~\onlinecite{CTLin2004, Voronkova93}. In order to keep these oxygen concentrations unchanged, the samples were detwinned at 400$^\circ$C under Ar flow. During detwinning, the samples were examined using a polarized light microscope.
Measurements of the superconducting transitions using a superconducting quantum interference device (SQUID) showed transition widths ($\Delta T_{c} < 2$ K), indicating good homogeneity of our samples. The hole doping level $p$ was determined from the known dependence of the out-of-plane lattice parameter $c$ and of $T_c$ on $p$~\cite{Liang_PRB2006}.
Polarized light microscopy and Laue x-–ray diffraction show that the $a$--, $b$-- and $c$--axes were always parallel to the edges of our rectangular-shaped single-domain crystals. The Raman spectra are reproducible at different spots of the $ab$-surface of each sample, further attesting to the homogeneity of the single crystals.
The in-plane orientation of the crystal was made using Laue diffraction, and further checked with the strong anisotropy of the Raman signal (chain related features appear indeed only when light is polarized along the $b$--axis~\cite{Slakey_PRB1989,Sacuto_SST} (see also Fig.~\ref{chains})).

\subsection{Raman measurements}
The Raman scattering experiments were performed in backscattering geometry on a Labram (Horiba Jobin-Yvon) single-grating Raman spectrometer equipped with a razor-edge filter. As incident photon wavelengths we used the $\lambda$=632.8 nm line of an He$^{+}$/Ne$^{+}$ mixed gas laser (red line), the $\lambda$=532 nm line of a Nd:YAG laser (green), the 514.52 nm of an Ar$^{+}$ laser (green) and the $\lambda$=488 nm of a diode-pumped solid state laser.
The incident laser power was $\sim$ 1 mW, and the beam focused through a $\times$50 long-working distance objective.
The samples were cooled down in a He-flow cryostat. Laser induced heating is always an issue in these kinds of measurements, and its amplitude can be hard to estimate at low temperatures where the anti-stokes signal is very low.
We could check that the heating was not larger than 5~K by looking at the well known superconductivity-induced renormalization of the 340 cm$^{-1}$ phonon mode~\cite{Macfarlane_PRB88,Friedl_PRL1990,Heyen_PRL1990,LeTacon_PRB2007,Altendorf_PRB1993,Bakr_PRB2009,Cardona_PhysicaC1999,Limonov_PRB2000, Misochko_SSC1994, Iliev_JRS1996,Panfilov_PRB97}. In all samples, the effect was observed at $T_c$, as expected (see also Fig.~\ref{B1gdop}). The spectral resolution of the spectrometer was about 2 cm$^{-1}$.

To describe the scattering geometries, we use the Porto notation A(BC)A', where A(B) and A'(C) stand for the propagation (polarization) directions of the incident and scattered light relative to the crystalline axes. All of the data presented in this paper were taken with the light propagation direction along the crystalline $c$--axis of our crystals, hence only the polarization orientations will be specified.
In the literature \cite{Friedl_PRL1990,Heyen_PRL1990,LeTacon_PRB2007,Altendorf_PRB1993,Bakr_PRB2009,Cardona_PhysicaC1999,Limonov_PRB2000}, the small orthorhombic distortion of YBa$_2$Cu$_3$O$_{6+x}$ is often neglected, and one refers to the phonons using the tetragonal (D$_{4h}$ point group) notation. Using the $XX$ and $YY$ polarizations for incident and scattered photons propagating along the crystallographic  $c$--axis, we couple to A$_{1g}$ and B$_{1g}$ symmetry phonons. In the more accurate orthorhombic notation (D$_{2h}$ point group), which will be used here~\cite{Macfarlane_PRB88,Liu_PRB88}, the modes have the A$_g$ symmetry.

\subsection{X-ray characterization of chain ordering}

\begin{figure}
\includegraphics[width=\linewidth]{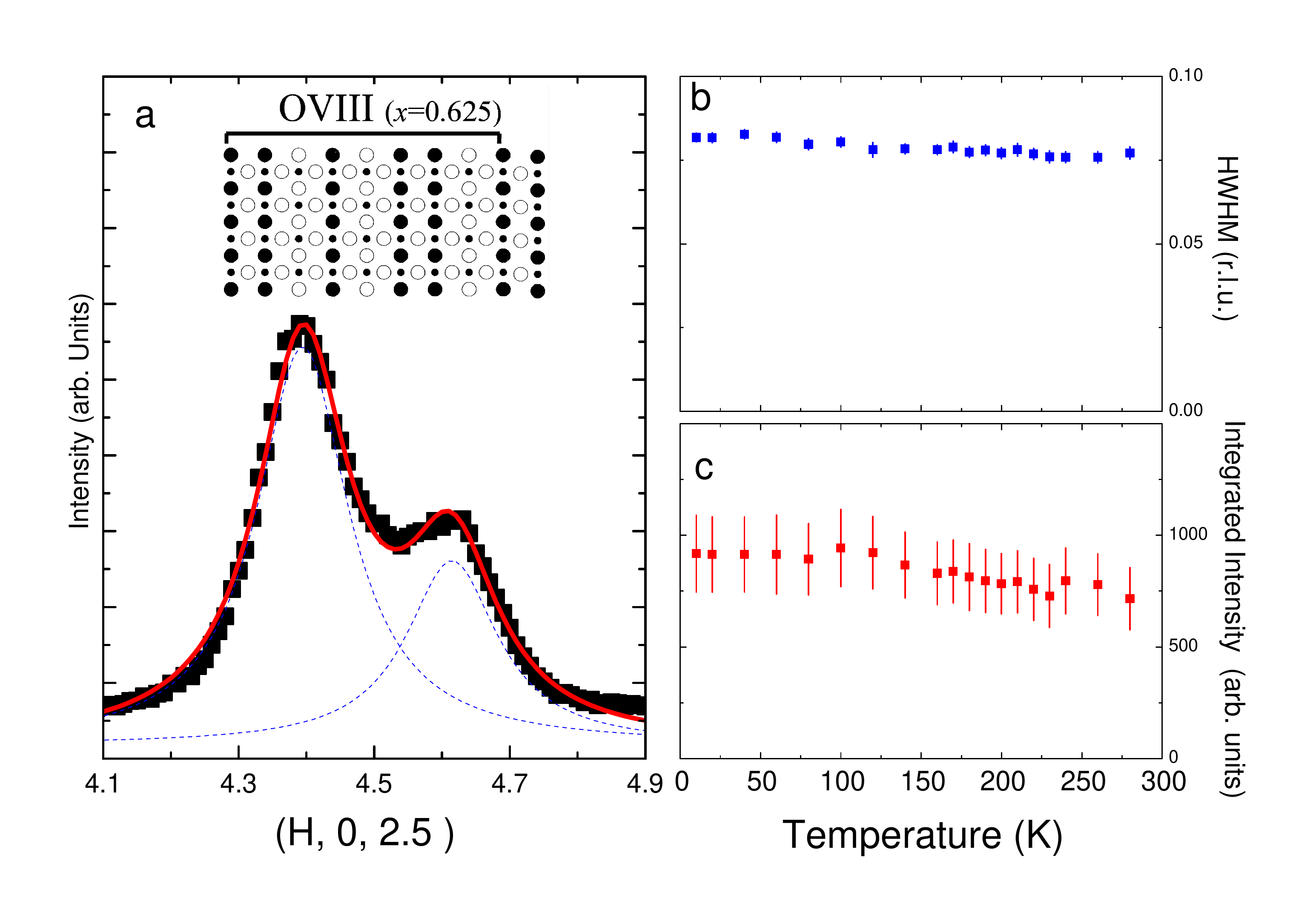}
\caption{a) X-ray diffuse scattering intensity around  the (4.5 0 2.5) reciprocal space point. A schematic of the ortho-VIII ordering is shown in the inset (from Ref.~\onlinecite{Andersen_PhysicaC99}). b) Temperature dependence of the half-width-half-maximum (HWHM) of the O-superstructure peaks. c) Temperature dependence of the integrated intensity of the O-superstructure peaks.} \label{xrays}
\end{figure}

X-ray diffuse scattering measurements were conducted at the high energy wiggler beamline BW5 at the Synchrotronstrahlungslabor (HASYLAB) at the Deutsches Elektronen-Synchrotron (DESY), with x-ray beam energy of 100 keV.
Following Ref.~\onlinecite{Strempfer_PRL2004}, hard x-ray diffuse scattering measurements on the YBa$_{2}$Cu$_{3}$O$_{6.6}$ crystal reveal broad structures centered at the (4.39, 0, 2.5) and (4.61, 0, 2.5) positions of reciprocal space (Fig.~\ref{xrays}). They correspond to an ortho-VIII type of oxygen superstructure with a correlation length of $\sim 16$~\AA, as expected for this oxygen content~\cite{Zimmermann_PRB2003}. A detailed mapping of the reciprocal space structure of the chains can be found in the supplementary information of Ref.~\onlinecite{Letacon_NatPhys2013}. The integrated intensities and widths are essentially temperature independent, in agreement with Ref.~\onlinecite{Strempfer_PRL2004}.
The YBa$_{2}$Cu$_{3}$O$_{6.75}$ and  YBa$_{2}$Cu$_{3}$O$_{6.55}$ single crystals were prepared with ortho-III and ortho-II order, respectively. The corresponding superstructure peaks were measured using soft x-ray resonant scattering \cite{Achkar_PRL2012, Blanco_PRL2013}, and correlation lengths of $\sim$ 37 \AA~for the ortho-III sample and $\sim$ 100 \AA~for the ortho-II one were found.

\section{Experimental Results}

\begin{figure}[b]
\includegraphics[width=\linewidth]{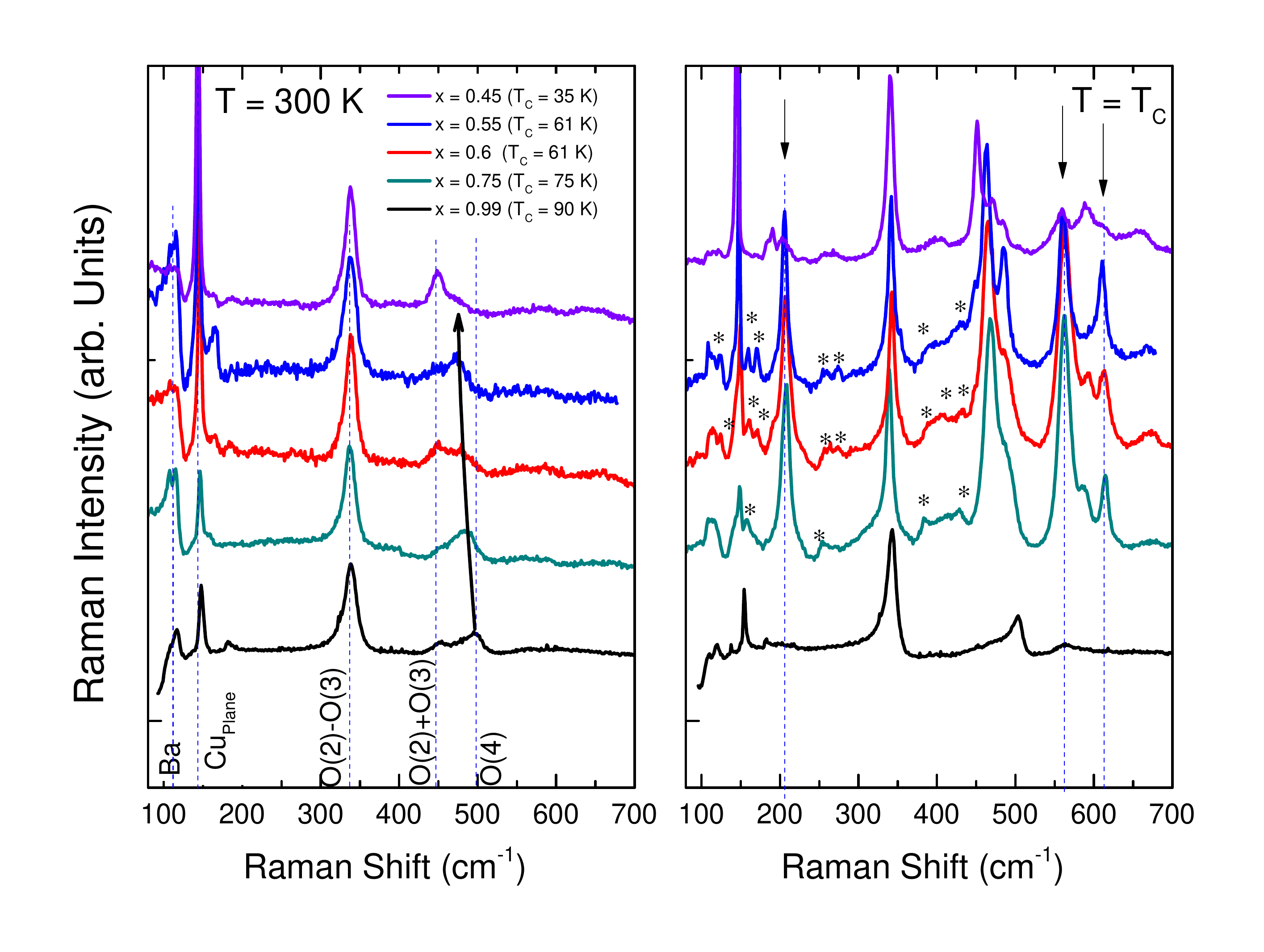}
\caption{Room temperature Raman spectra of detwinned optimally doped YBa$_{2}$Cu$_{3}$O$_{6.99}$ and underdoped YBa$_{2}$Cu$_{3}$O$_{6.75}$, YBa$_{2}$Cu$_{3}$O$_{6.6}$, YBa$_{2}$Cu$_{3}$O$_{6.55}$ and YBa$_{2}$Cu$_{3}$O$_{6.45}$ single crystals in the $XX$ channel taken with an Neon laser line ($\lambda$ =632.8 nm) at room temperature and $T = T_c$. The mode assignment corresponds to Refs.~\onlinecite{Cardona_PhysicaC1999}. For clarity, the absolute intensity of the spectra were normalized and a vertical offset has been added.
} \label{RTdata}
\end{figure}

We have performed temperature dependent Raman scattering measurements on all the samples.
Room temperature Raman spectra for the five single crystals are shown on the left panel of Fig.~\ref{RTdata}. They consist of a broad, featureless electronic continuum on top of which are superimposed sharp optical phonons. The five Raman active modes associated with the ortho-I unit cell (full CuO chains) are visible at room temperature in all samples. These modes have been widely studied in the literature~\cite{Macfarlane_PRB88,Slakey_PRB1989,Sacuto_SST,Burns_SSCMcCarty_PRB1990,Friedl_PRL1990,Heyen_PRL1990,LeTacon_PRB2007,Altendorf_PRB1993,Bakr_PRB2009,Cardona_PhysicaC1999,Limonov_PRB2000,Liu_PRB88, Misochko_SSC1994, Iliev_JRS1996,Panfilov_PRB97}. The corresponding atomic displacement as well as the typical energies corresponding to these modes are given in table~\ref{tab: param}.
Note that these values are strongly doping dependent~\cite{Macfarlane_PRB88,Altendorf_PRB1993}. In particular, the mode close to 500 cm$^{-1}$ mode softens a lot with diminishing oxygen content.

\begin{table}[t]
\begin{ruledtabular}
\begin{tabular}{|c|c|}
displacement & Approx. Energy (cm $^{-1}$)\\
\hline
Ba($z$) & 115\\
\hline
mixed Ba/Cu(2)($z$) & 150\\
\hline
O(2)-O(3)($z$) & 340\\
\hline
O(2)+O(3)($z$) & 435\\
\hline
O(4)($z$) & 500\\
\end{tabular}
\end{ruledtabular}
\caption{\label{tab: param} Typical phonon (approximate) energies (in cm$^{-1}$) for the Raman allowed modes of the ortho-I structure.}
\end{table}

\begin{figure}
\includegraphics[width=\linewidth]{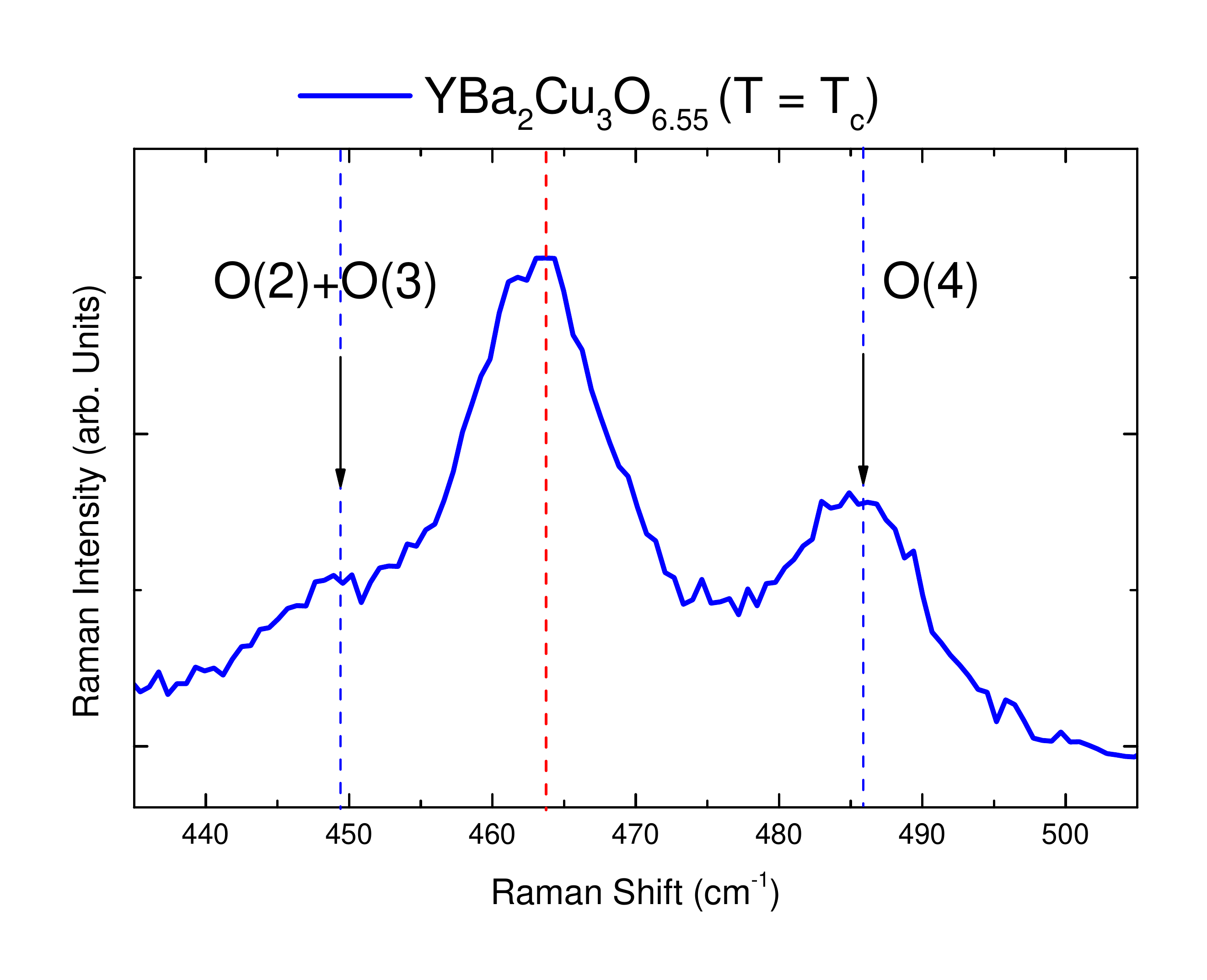}
\caption{Detailed view of the XX Raman spectrum of YBa$_{2}$Cu$_{3}$O$_{6.55}$ at $T=T_c$} \label{o2zoom}
\end{figure}

The main result of this work is clearly visible in the raw Raman data at temperature $T=T_c$ (right panel of Fig.~\ref{RTdata}). At optimal doping, no dramatic changes in the spectra are seen upon cooling, other than the expected sharpening of the aforementioned modes. In all underdoped compounds, on the other hand, the low temperature spectra appear quite different from the room temperature ones. In addition to many weak features, which are sample dependent and related to superstructures induced by O-dopants, we observe a very large scattering intensity around 205, 465, 560 and 610 cm$^{-1}$.
In  YBa$_{2}$Cu$_{3}$O$_{6.6}$ and  YBa$_{2}$Cu$_{3}$O$_{6.75}$, the peak close to 465 cm$^{-1}$ overlaps with the in-phase vibration of the planar oxygen atoms (labeled '$O(2)+O(3)$' in table~\ref{tab: param}) and the apical O(4) mode. A detailed look to the data of YBa$_{2}$Cu$_{3}$O$_{6.55}$ (Fig.~\ref{o2zoom}), which is more ordered than the other underdoped compounds, and has therefore sharper phonons, demonstrates that the 465 cm$^{-1}$  feature is distinct from the two surrounding phonons.


\subsection{Raman selection rules}
\begin{figure}
\includegraphics[width=\linewidth]{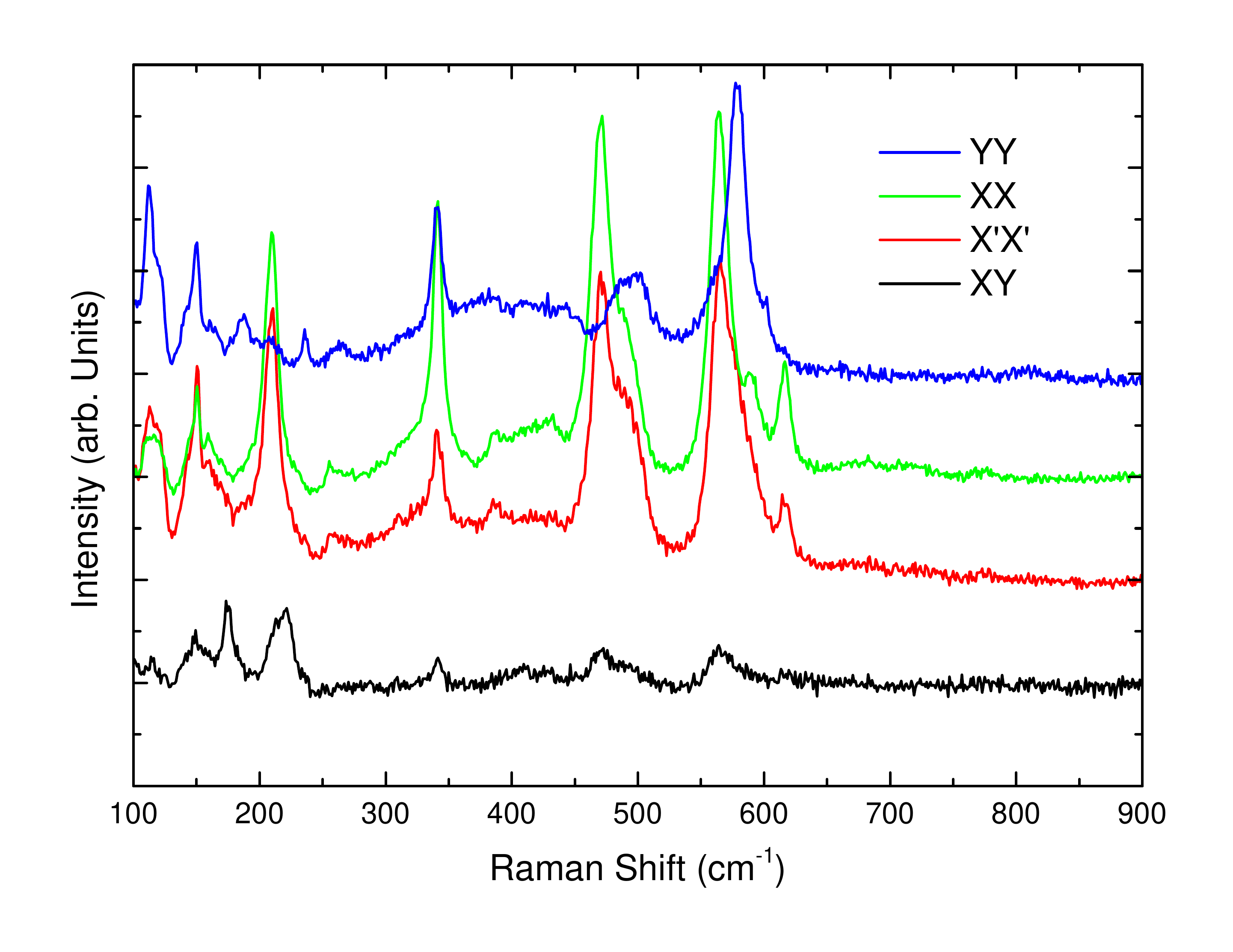}
\caption{Raman spectra of YBa$_{2}$Cu$_{3}$O$_{6.75}$ obtained in various scattering geometries at $T = T_c$. Spectra intensities have been normalized at 900 cm$^{-1}$ and shifted vertically for clarity.} \label{selection}
\end{figure}

In Fig.~\ref{selection}, we have plotted the Raman intensity measured in various scattering geometries at $T=T_c$ for the YBa$_{2}$Cu$_{3}$O$_{6.75}$ sample.
The new modes are visible only in the $XX$ and $X^{\prime}X^{\prime}$ scattering geometries which select respectively the $A_{g}$ and $A_{g}+B_{1g}$ modes (in the orthorhombic notation). This shows that these features have the $A_{g}$ symmetry.

\subsection{Role of the chains}
\label{sectionchains}
In the 123 family, the doping level is controlled by the occupation of O positions in the CuO chains, which can lead to the formation of superstructures with different periodicity along the $a$--axis, depending on the heat treatment and on the oxygen content~\cite{Andersen_PhysicaC99,Zimmermann_PRB2003}. As explained in the Experimental Details section, these superstructures have been characterized, and can give rise to a large number of new Raman active modes (different for each types of unit cell). In the ortho-II structure observed for oxygen contents close to $x \sim 0.5$, where empty and full chains alternate with correlation lengths larger than 100~\AA, the superlattice induced modes (6 new Raman active phonons with the A$_g$ symmetry) are visible, but remain considerably weaker than those of the ortho-I allowed modes at all temperatures ~\cite{Iliev_PRB2008,Misochko_SSC1994, Iliev_JRS1996}.
In view of the decreasing correlation length of the superstructures (37 \AA~for ortho-III and only 16 \AA~for ortho-VIII), it is reasonable to expect the related phonons to be weak and sample dependent, such as those marked with ``$\ast$'' in Fig.~\ref{RTdata}-b.

The strong features around 205, 465, 560, and 610 cm$^{-1}$ observed in these three samples do not meet this expectation. Further, at variance with the weaker modes mentioned above, their energies are essentially doping independent, suggesting a common physical origin. In the most underdoped sample investigated here, YBa$_{2}$Cu$_{3}$O$_{6.45}$, these modes are also observed but remain extremely weak.

\begin{figure}
\includegraphics[width=\linewidth]{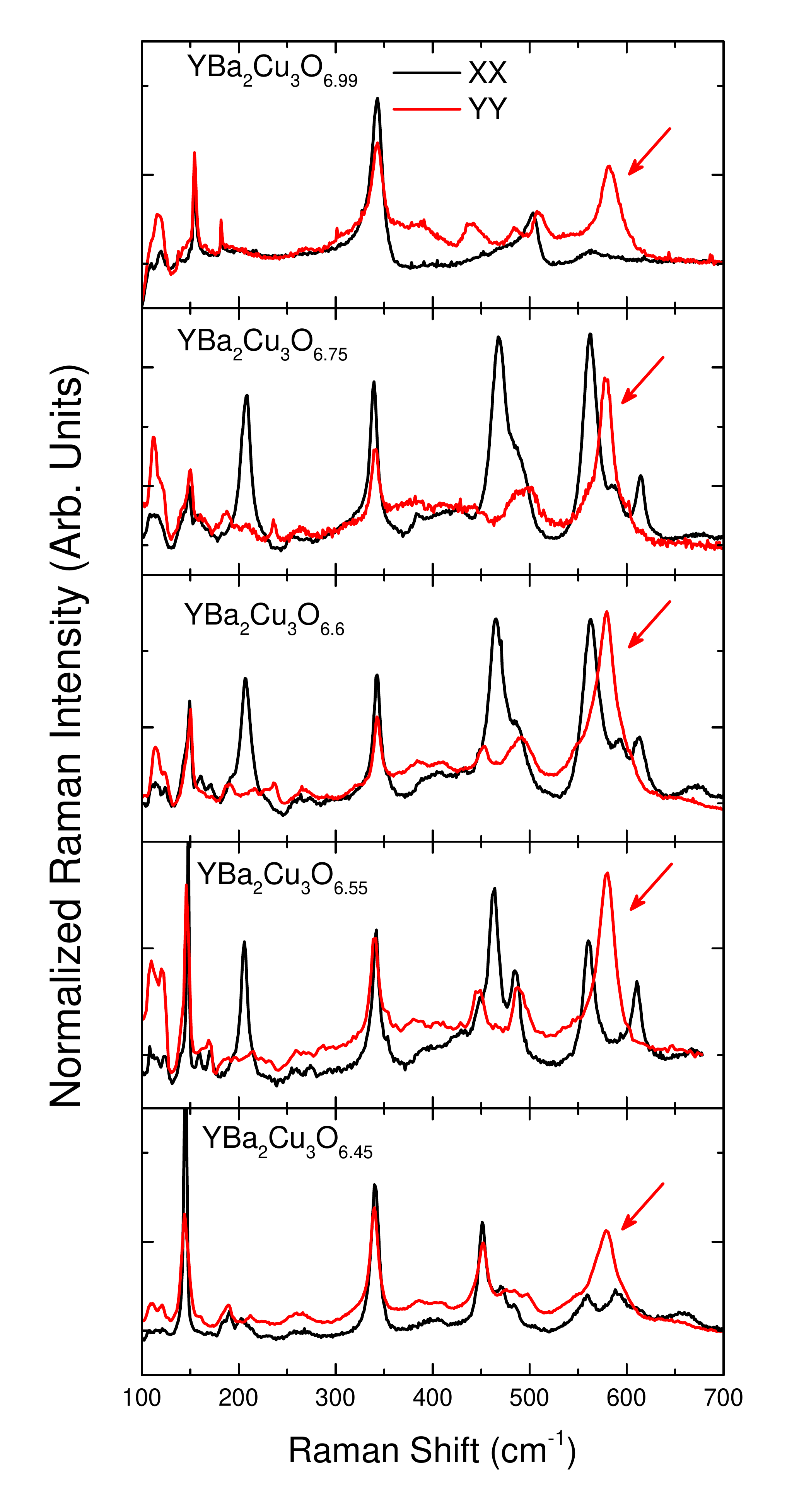}
\caption{Comparison of the Raman spectra obtained at $T = T_c$ in XX and YY scattering geometries. The arrows indicate the strongly temperature dependent mode at 590 cm$^{-1}$, associated with the O-vibration within broken CuO chains.} \label{chains}
\end{figure}

A look at the data taken in the YY geometry (which also selects the $A_{g}$ symmetry) provides further evidence against an O-structure related origin for these modes. In this geometry, the new modes are hardly visible. The observation of the 560, and 610 cm$^{-1}$ features is prohibited by the presence of additional phonons arising from the broken CuO chains~\cite{Liu_PRB88,Burns_SSC,Panfilov_PRB97,Wake_PRL1991}, in particular the one at 590 cm$^{-1}$, which is very intense for the 632.8 nm incident wavelength and strongly temperature dependent at all doping levels (\textit{i.e.} also at the optimal doping where the new modes are not seen). At 205 cm$^-{1}$, where chain-related features are not present, we do not observe any peak.

It is important to emphasize that in our detwinned single crystals, the chain related structures are visible \textit{only} in the YY geometry and are seen at all doping levels (Fig.~\ref{chains}). Other effects such as laser induced local disorder of the oxygen structures~\cite{Iliev_JRS1996,Bock_PRB1995} are also expected to occur only in the YY channel, albeit for much higher power densities than the one used here. Those would furthermore also occur in the optimally doped compound where no effect is seen here. This allows us to rule out already at this stage any chain-related origin for the 205, 465, 560 and 610 cm$^{-1}$ modes. This conclusion will be reinforced by looking at resonance effects (section~\ref{sectionresonance}) and at the detailed temperature dependence of these modes (section~\ref{sectiontdep}).

\subsection{ZZ and XZ geometries}

\begin{figure}
\includegraphics[width=\linewidth]{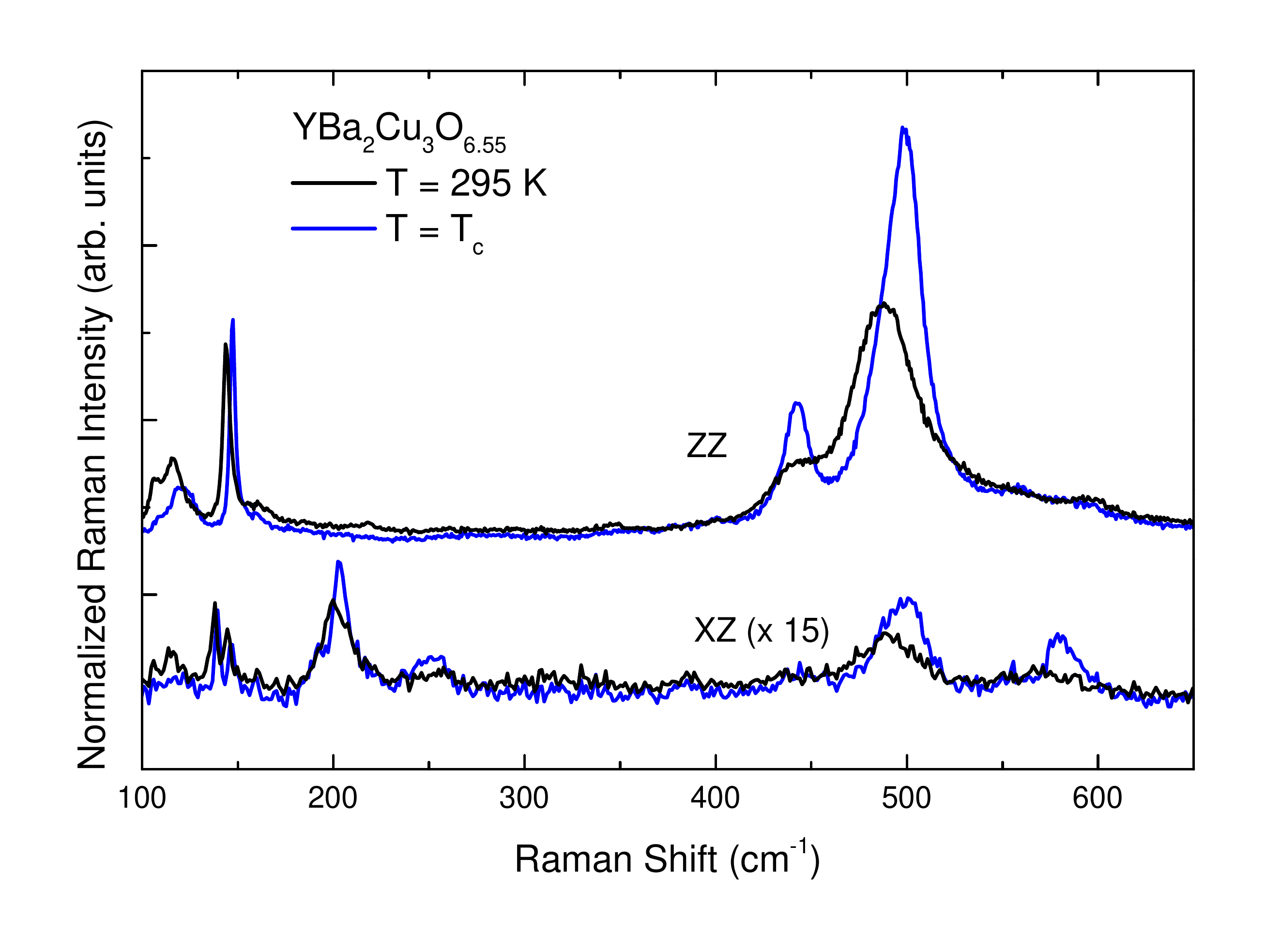}
\caption{Comparison of the Raman spectra obtained at $T = T_c$ and room temperature in XZ and ZZ scattering geometries on  YBa$_{2}$Cu$_{3}$O$_{6.55}$. For clarity, the XZ spectra has been magnified 15 times and a vertical offset has been applied between the two sets of data.} \label{caxis}
\end{figure}

Before addressing these points, we present data taken with the light polarized in the $ac$ plane on the ortho-II sample YBa$_{2}$Cu$_{3}$O$_{6.55}$.
On Fig.~\ref{caxis} we show the spectra measured in XZ and ZZ geometries that respectively probe the $B_{2g}$ and $A_g$ geometries at room temperature and $T = T_c$, and that agree well with previous reports~\cite{McCarty_PRB1990,Altendorf_PRB1993,Iliev_PRB2008}.
At variance with the data obtained using in-plane scattering geometries, no extra features were observed at low temperature in these spectra.
We note that in the $B_{2g}$ spectra, a Raman active phonon is seen at 203 cm$^{-1}$ and previously attributed to planar vibrations of the apical $O(4)$~\cite{McCarty_PRB1990}. The energy of this mode is very close from the one of the feature we observe at low temperature around 205 cm$^{-1}$. Unlike this feature however, the weak B$_{2g}$ peak is already seen at room temperature and appears to have a conventional temperature dependence. Furthermore, since none of the other new modes seen in the XX geometry at 465, 560, and 610 cm$^{-1}$ appear in this spectra, we are led to conclude that the energy proximity between the 205 cm$^{-1}$ feature and the $B{_2g}$ mode is very likely coincidental.
Note finally that the absence of a strongly temperature dependent feature at 600 cm$^{-1}$ in the $ZZ$ data allow us to rule out laser-induced damages of the chains during the measurement, as reported in ref.~\onlinecite{Burns_SSC}.
Together with our previous observations, this indicates that the $A_g$ Raman tensor associated with the new modes is highly anisotropic ($\alpha_{XX} >> \alpha_{YY}, \alpha_{ZZ}$).

\subsection{Resonance effects}
\label{sectionresonance}
\begin{figure}[t]
\includegraphics[width=0.95\linewidth]{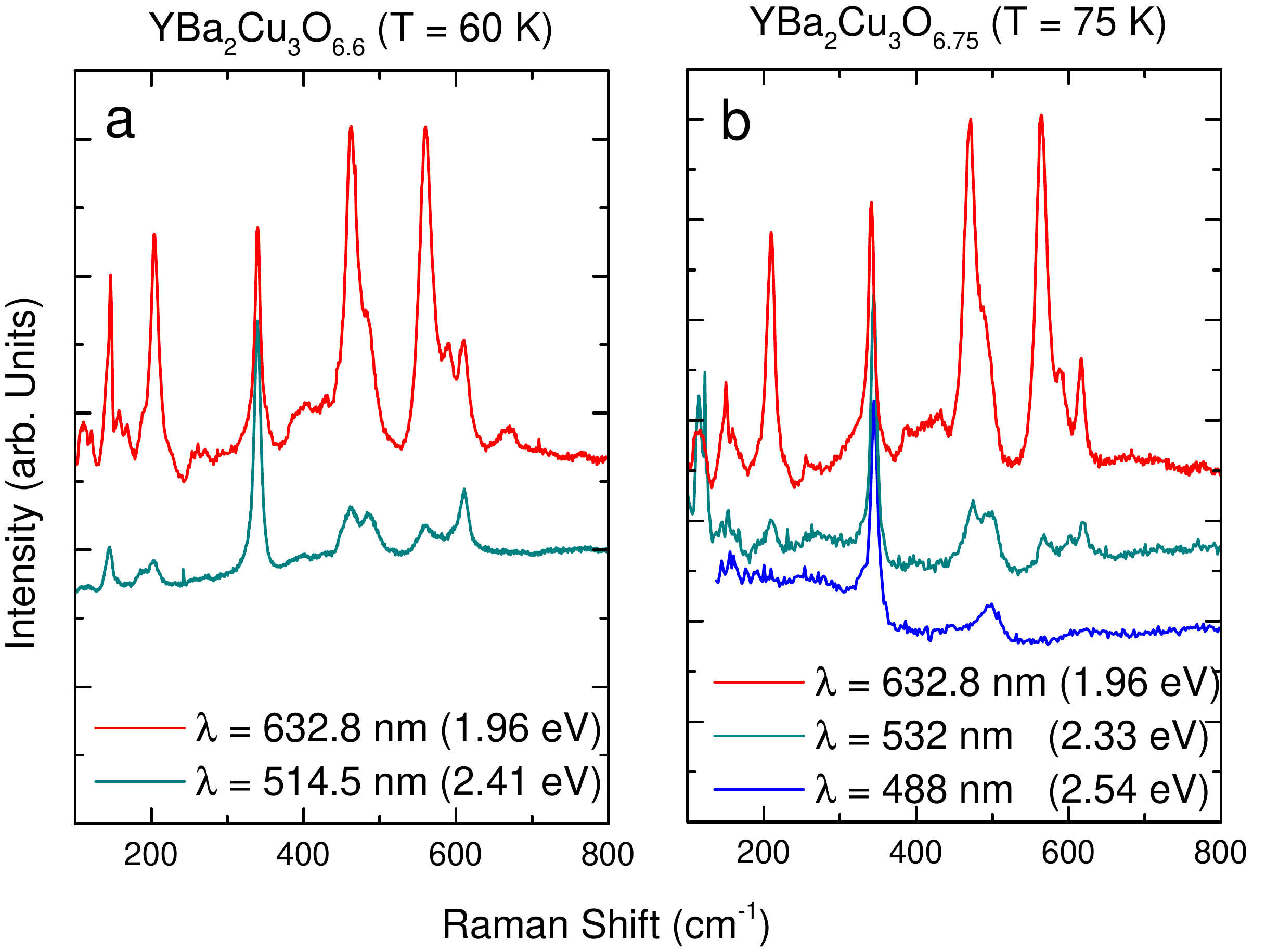}
\caption{a) XX Raman spectra of YBa$_{2}$Cu$_{3}$O$_{6.6}$ measured at $T =$ 60 K with red ($\lambda$ = 632.8 nm) and green ($\lambda$ = 514.5 nm) incident laser lights. a) XX Raman spectra of YBa$_{2}$Cu$_{3}$O$_{6.75}$ measured at $T =$ 75 K with red ($\lambda$ = 632.8 nm), green ($\lambda$ = 532 nm) and blue ($\lambda$ = 488 nm) incident laser lights. For clarity, the intensities of the spectra have been normalized to the one of the 340 cm$^{-1}$ phonon.}
\label{resonance}
\end{figure}

From now on, we will focus on the spectra obtained in the XX scattering geometry where the chain related complications are absent. Having determined the Raman selection rules for the new modes, we now turn to their resonance properties, that is, the evolution of their intensity as function of the incident laser light energy. The resonance profile of an excitation can bring useful information as it can be directly related to the intermediate state involved in the scattering process and has been widely used to study selectively electronic, phononic and magnetic excitations in cuprates~\cite{Cooper_PRB1993,Heyen_PRL1990,Panfilov_PRB97,Rubhausen_PRB96,Blumberg_PRB1994,Kang_PRL96,LeTacon_PRB05,Le Tacon_NaturePhysics06,Li_PRL2013}.
Interestingly, when using green incident laser light, the new modes are still clearly visible, but their relative intensity is strongly reduced compared to \textit{e.g.} the 340 cm$^{-1}$ phonon (Fig.~\ref{resonance}). Using the blue line, the modes disappear completely. This trend differs from previous reports concerning the other Raman active modes~\cite{Heyen_PRL1990}, and from the behavior of the chain defect induced features~\cite{Burns_SSC,Panfilov_PRB97, Iliev_JRS1996,Wake_PRL1991}, whose intensities are maximized for green and yellow incident lights, respectively.

Since most of the phonon studies carried out on underdoped YBa$_{2}$Cu$_{3}$O$_{6+x}$ have used green light~\cite{Macfarlane_PRB88, Misochko_SSC1994, Iliev_JRS1996,Altendorf_PRB1993,Limonov_PRB2000}, this might explain why this effect has not been reported by previous Raman studies. We note, however, that the 205, 560 and 610 cm$^{-1}$ features were in fact seen in the XX channel in the data presented in ref.~\onlinecite{Iliev_PRB2008} in an ortho-II crystal with  $T_c$ = 57.5 K (the effect at 80 K is weaker than the one reported here - likely due to a lower doping level - but still clearly visible) as well as in a sample with $T_c$ = 70 K measured with the 647.1 nm Kr$^+$ line at 15 K in Ref.~\onlinecite{Panfilov_PRB97}, where they are referred to as ``weak'' (which is indeed the case at low temperature) and were not further studied.

\subsection{Detailed temperature dependence}
\label{sectiontdep}

\begin{figure}
\includegraphics[width=\linewidth]{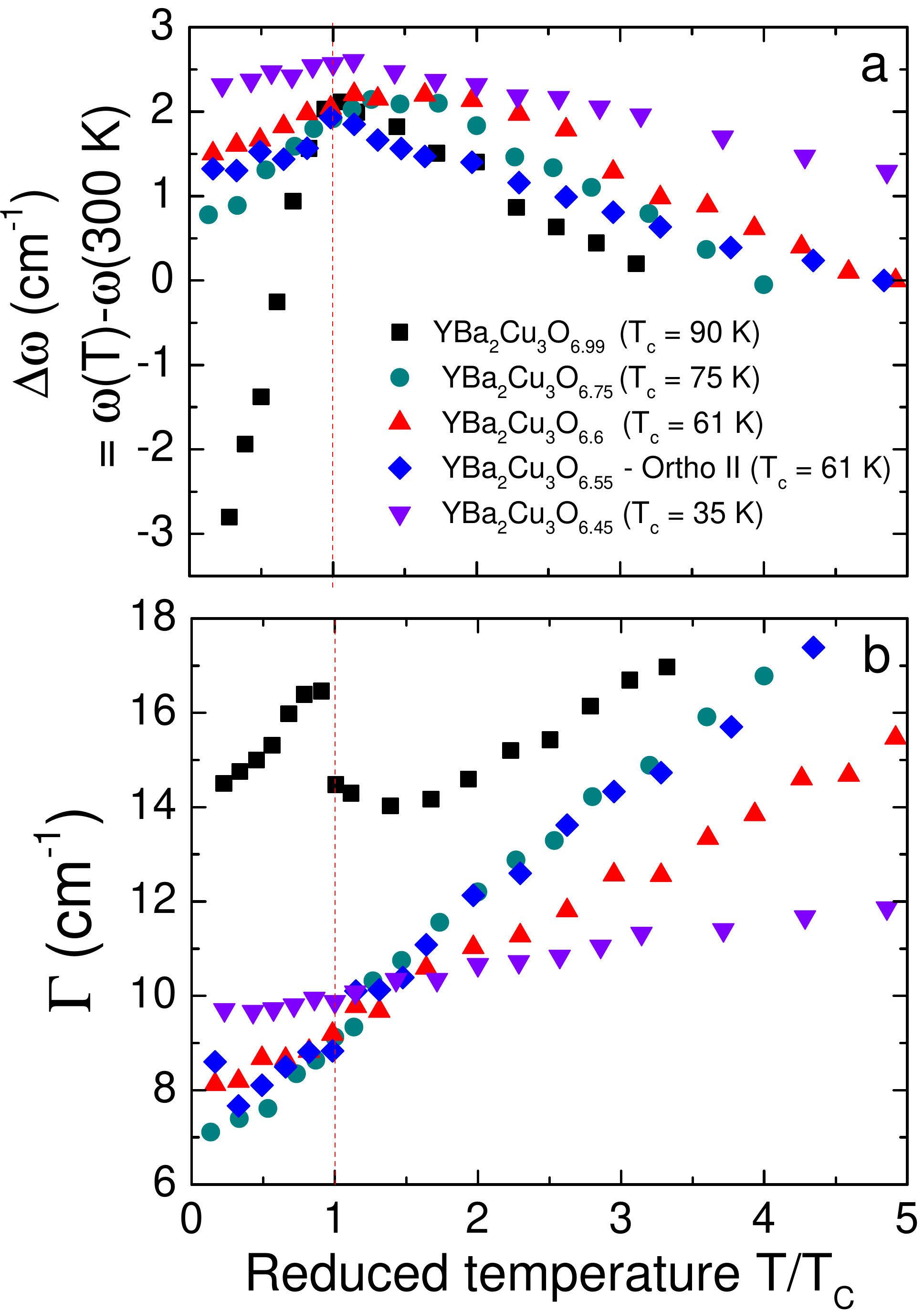}
\caption{Renormalization of the 340 cm$^{-1}$ 'buckling' mode through $T_c$ for the different doping levels.}
\label{B1gdop}
\end{figure}

\begin{figure*}
\includegraphics[width=\linewidth]{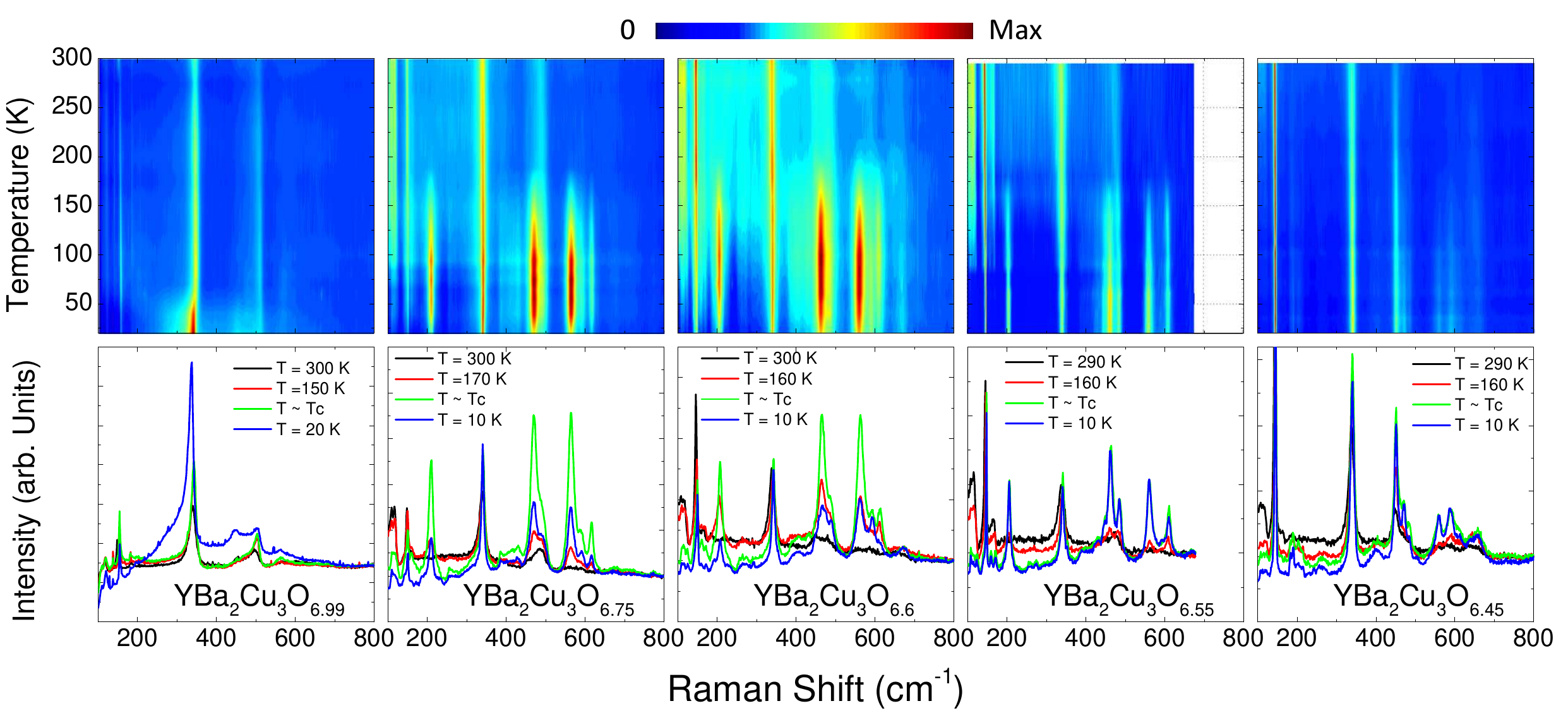}
\caption{Upper panel: Colormap representation of the temperature dependence of the Raman intensity measured with $\lambda$=632.8 nm in the XX channel for optimally doped YBa$_{2}$Cu$_{3}$O$_{6.99}$, and underdoped  YBa$_{2}$Cu$_{3}$O$_{6.75}$, YBa$_{2}$Cu$_{3}$O$_{6.6}$, YBa$_{2}$Cu$_{3}$O$_{6.55}$, and YBa$_{2}$Cu$_{3}$O$_{6.45}$. Lower panel: Raman spectra of the same compounds for selected temperatures.
} \label{maps}
\end{figure*}

We now take a closer look at the temperature dependence of the Raman spectra shown in Fig.~\ref{maps}.
At optimal doping, a redistribution of the electronic continuum as well as a renormalization of the lineshape of those phonons that couple strongly to the electrons (including the 340~cm$^{-1}$ buckling mode and the low frequency c-axis polarized Ba and Cu modes) are seen upon cooling below $T_c$ (see Fig.~\ref{B1gdop}).
This is a direct consequence of the opening of the superconducting gap in the electronic density of states, an effect widely studied and quantitatively well understood~\cite{Altendorf_PRB1993,Cardona_PhysicaC1999,Friedl_PRL1990,Limonov_PRB2000,Schnyder_PRB2007,Bakr_PRB2009}. In the underdoped compounds, the amplitude of the superconductivity-induced renormalization is strongly reduced, in agreement with previous studies~\cite{Altendorf_PRB1993,Limonov_PRB2000}. This doping evolution is understood as a consequence of the increase of the maximal gap amplitude as doping is reduced, and of the opening of the normal state pseudogap that suppresses the electronic density of states (and consequently the electron-phonon coupling) at the Fermi level.

\begin{figure}
\includegraphics[width=0.9\linewidth]{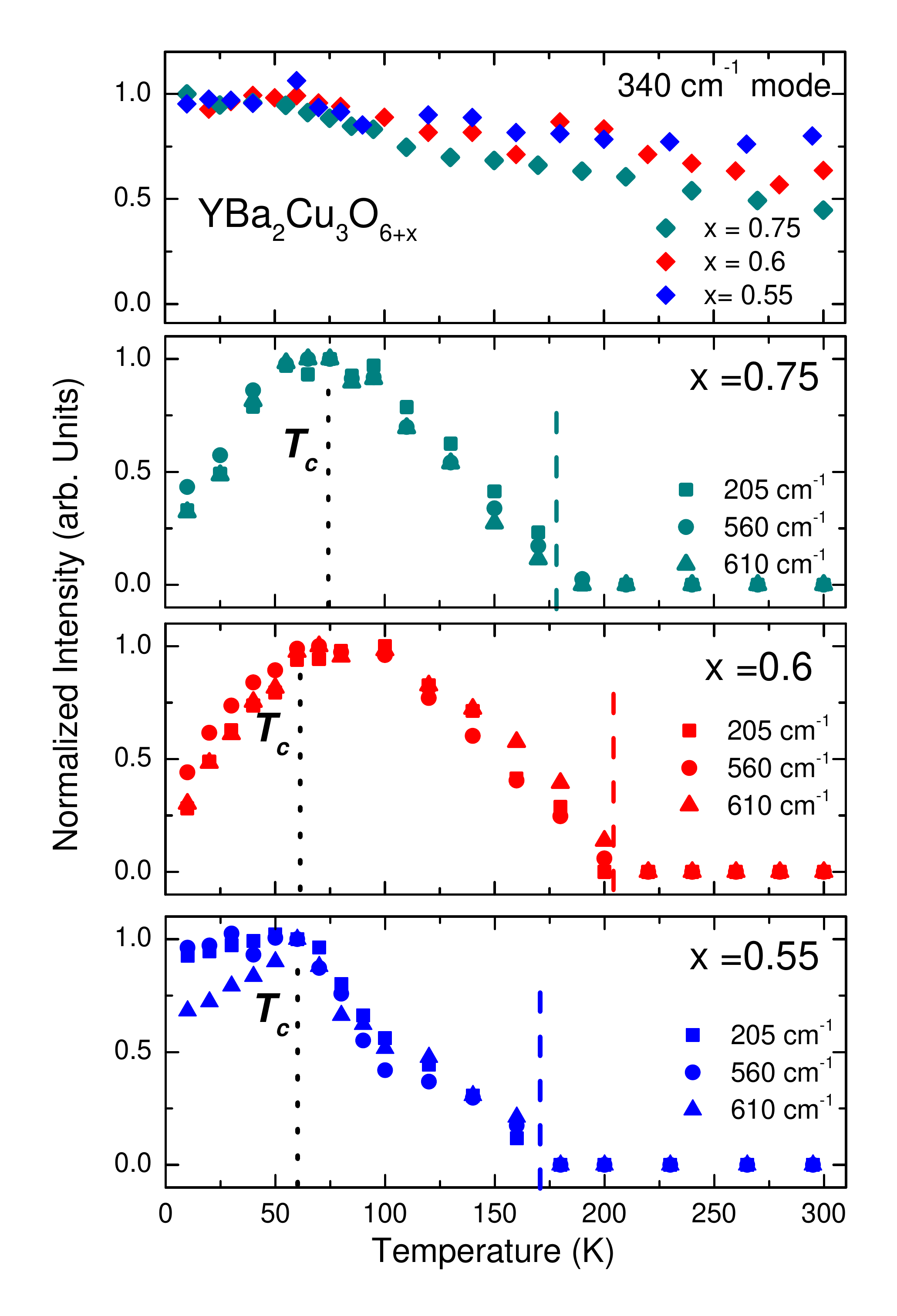}
\caption{Temperature dependence of the integrated intensity of the modes at 340~cm$^{-1}$ (upper panel) and 205, 560 and 610~cm$^{-1}$ (lower panel) in  YBa$_{2}$Cu$_{3}$O$_{6.75}$ samples (green symbols), YBa$_{2}$Cu$_{3}$O$_{6.6}$ (red symbols), and YBa$_{2}$Cu$_{3}$O$_{6.55}$ samples (blue symbols).}
\label{Tdep_phonons}
\end{figure}

The intensity maps of Fig.~\ref{maps} show clearly that the new modes in the underdoped YBa$_{2}$Cu$_{3}$O$_{6.75}$, YBa$_{2}$Cu$_{3}$O$_{6.6}$ and YBa$_{2}$Cu$_{3}$O$_{6.55}$ samples appear in the normal state, far above their respective superconducting transition temperatures.
The integrated intensities of the 205, 560 and 610 cm$^{-1}$ modes are displayed as a function of temperature in Fig.~\ref{Tdep_phonons}. We leave the feature around 465 cm$^{-1}$ aside, because as mentioned earlier we cannot exclude that part of its intensity originates from the O(2)+O(3) and/or the O(4) Raman active modes (though it is clearly distinct from them, as evidenced in the ortho-II samples where all three features can be resolved as seen in Fig.~\ref{o2zoom}).
In a given sample, the three modes behave in a very similar manner, their intensity being maximized around $T_c$. A significant decrease of the intensity in the superconducting state is seen especially in YBa$_{2}$Cu$_{3}$O$_{6.75}$ and YBa$_{2}$Cu$_{3}$O$_{6.6}$.
The sensitivity of the measurement further allows us to study the doping dependence of the onset temperature, which appears non-monotonic and reaches its maximum value in YBa$_{2}$Cu$_{3}$O$_{6.6}$ around 200 K. In both  YBa$_{2}$Cu$_{3}$O$_{6.75}$ and YBa$_{2}$Cu$_{3}$O$_{6.55}$ the onset temperatures are around 175 K.
The frequencies and linewidths of the new modes (not shown here) do not display any remarkable behavior. The modes harden and narrow slightly upon cooling, as expected from anharmonicity.

\section{Discussion}
We start our discussion with a brief summary of our main experimental findings. We have shown that new, intense phonon features appear below $\sim$ 200 K in the polarized Raman spectra of three moderately underdoped YBa$_2$Cu$_3$O$_{6+x}$ single crystals ($x=0.55$, 0.6 and 0.75). They could not be detected at optimal doping ($x=0.95$), and become extremely weak at lower doping ($x=0.45$). These new features are best seen with red incident laser light, in contrast to regular Raman-active phonons (which are best seen with green light~\cite{Heyen_PRL1990}) and chain-induced defect modes (which are best seen with yellow light~\cite{Wake_PRL1991,Panfilov_PRB97}). The intensity of these new features is maximal at the superconducting transition temperature, and decreases upon cooling below $T_c$. The polarization selection rules, the resonance conditions, and the doping and temperature dependencies of these modes allow us to rule out a chain related origin.

The number of phonon branches is determined by the lattice symmetry. It cannot change with temperature in the absence of a structural phase transition that (i) lifts the degeneracy of some of the modes, (ii) breaks the IR/Raman selection rules (when suppressing the inversion symmetry of the crystal or driving an atom out of an inversion-symmetric site) or (iii) back-folds the phonon dispersion and thus give rises to new optical modes at the $\Gamma$ point (when increasing the size of the original unit cell). The first two cases can easily be ruled out in our case. Clearly the new features we observe do not originate from splitting of the Raman active phonons of the YBa$_{2}$Cu$_{3}$O$_7$ structure; the modes that we are probing in the XX geometry are actually non-degenerate A$_g$ modes. The breakdown of inversion symmetry seems also unlikely as the new feature energies do not match those of known IR active modes~\cite{Bohnen_EPL2003,Bernhard_2002}.
This therefore indicates that we are in the third situation.

The temperature and doping dependencies of the new modes, and in particular the fact that their intensity is maximized at $T_c$, is strongly reminiscent of the recently discovered CDW correlations in underdoped YBa$_{2}$Cu$_{3}$O$_{6+x}$~\cite{Ghiringhelli_Science2012,Achkar_PRL2012,Chang_NaturePhysics2012}, which compete with superconductivity. Studies carried out on the same single crystals using resonant x-ray scattering, unveiled a biaxial CDW in YBa$_{2}$Cu$_{3}$O$_{6.6}$~\cite{Ghiringhelli_Science2012} and YBa$_{2}$Cu$_{3}$O$_{6.75}$~\cite{Achkar_PRL2012} single crystals, and a more anisotropic (but still bi-axial) signal in ortho-II YBa$_{2}$Cu$_{3}$O$_{6.55}$~\cite{Blackburn_PRL2013,Blanco_PRL2013}.
Qualitatively, it is natural to associate the appearance of new optical modes to the folding of the phonon dispersion in the new BZ as the CDW appears.
The CDW incommensurability is only slightly doping dependent, decreasing from $\delta \sim$ 0.32 to 0.3 (along the $b^*$ direction) when going from  $x=0.55$~\cite{Blackburn_PRL2013} to $x=0.75$~\cite{Blackburn_PRL2013,Achkar_PRL2012}. This is consistent with the doping independence of the mode frequencies.
The appearance of new phonons in an electronically-driven charge ordered state has been reported for other materials such as transition metal dichalcogenides (see \textit{e.g.} Ref. ~\onlinecite{Sugai_PSSB1985} and references therein), manganites~\cite{Yoon_PRL2000} but also in stripe-ordered nickelates~\cite{Gnezdilov_JLT2005, Blumberg_PRL1997}. When the CDW is commensurate with the lattice, these effects are strong and can be quantitatively understood based on group symmetry analysis; the new zone center modes are simply the zone-boundary phonons of the high temperature phase. In the case of incommensurate structures, or in the fluctuating regime above the phase transition, these effects are usually much weaker.

In the 123 system, the CDW-induced atomic displacements are incommensurate with the underlying lattice, and the intensity ratio of the CDW satellite reflections and neighboring Bragg reflections yields an estimate of $\sim 10^{-3} a$ for their amplitude~\cite{Chang_NaturePhysics2012}. Moreover, the CDW state always remains short-range ordered, at least in the absence of a high magnetic field~\cite{LeBoeuf_NaturePhysics2013,Shekhter_Nature2013,Meier_PRB 2013}. In this sense, the amplitude of the effect reported here appears rather strong. One should, however, keep in mind the strongly resonant character of the new modes, \textit{i.e.} the strong dependence of their Raman intensity on the incident photon frequency. Among the wavelengths investigated here, we observed maximal intensity for red photons with energy 1.96 eV, which indicates that the intermediate states for Raman scattering in this photon energy range (close to the charge transfer gap energy between the O(2p) and Cu(3d) levels~\cite{Cooper_PRB1993}) are extremely sensitive to the CDW fluctuations.

It would be interesting to assign the new modes to atomic displacement patterns. While
this is relatively straightforward for commensurate CDWs, the task is far more complex here owing to the incommensurate nature of the modulation. In addition, our recent x-ray diffuse scattering reciprocal space mapping revealed a complicated structure factor~\cite{Letacon_NatPhys2013}, and a clear real-space picture of the atomic displacements associated with this modulated state has not yet emerged. Finally, we recall that in the presence of oxygen superstructures in the underdoped compound, the original phonon dispersion is still poorly known to-date. The only reasonable starting point is the phonon spectrum of fully oxygenated YBa$_2$Cu$_3$O$_7$ that has been widely studied using INS and IXS~\cite{Pintschovius_PRB2004,Pintschovius_PRL2002,Reznik_Nature2006,Raichle_PRL2011,Baron_JPCS2008,Bohnen_EPL2003}. The best candidates for the 560 and 610 cm$^{-1}$ features are the branches associated with in-plane vibrations of the oxygen atoms in CuO$_2$ planes. These branches comprise bond-stretching modes previously argued to couple dynamically with charge inhomogeneities in the CuO$_2$ planes~\cite{Pintschovius_PRB2004,Pintschovius_PRL2002,Reznik_Nature2006}.
The assignment of the 205 cm$^{-1}$ feature is more subtle since it is located in a denser region of the phonon spectrum. The energetically closest mode experimentally measured so far is associated with in-plane motion of Y and planar Cu atoms (IR-active at the zone center)~\cite{Bohnen_EPL2003}.
Finally, the 465 cm$^{-1}$ peak might originate from the backfolding of the branch dispersing from the O(2)+O(3) Raman active mode~\cite{Bohnen_EPL2003}.

\begin{figure}
\includegraphics[width=0.95\linewidth]{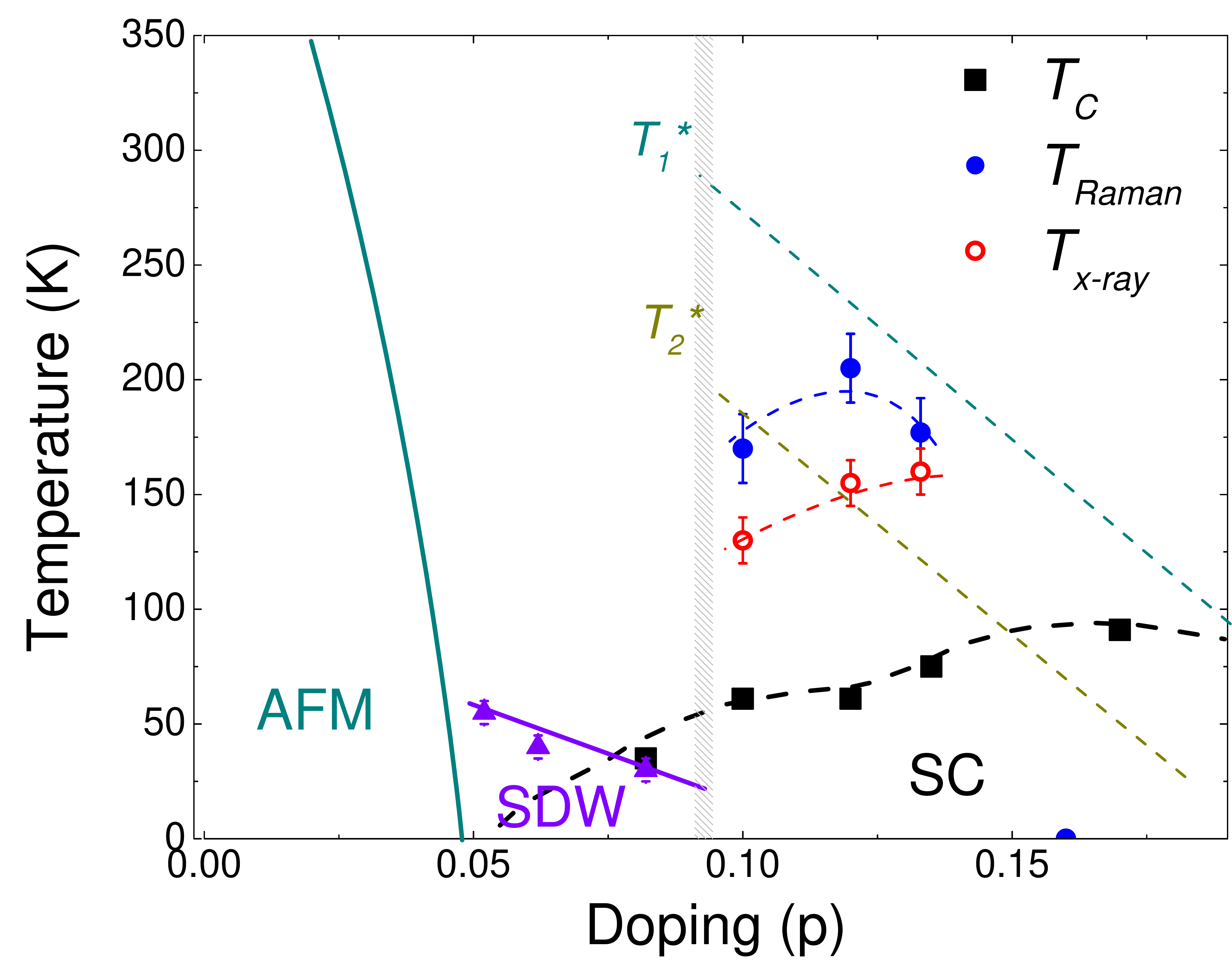}
\caption{Schematic phase diagram for the Y123 family. $T_{Raman}$ corresponds to the appearance temperature of the new modes at 205, 560 and 610 cm$^{-1}$, while $T_{x-ray}$ stands for the inset of the CDW peak seen in soft x-ray scattering experiments~\cite{Achkar_PRL2012,Ghiringhelli_Science2012,Blanco_PRL2013} in the same samples. Other lines are discussed in the text.}
\label{PhaseDiagram}
\end{figure}

We now turn to the temperature dependence of the new Raman features, and more specifically to their onset temperature. The three samples in which the Raman phonons are the strongest are those in which a superstructure peak has been clearly identified using soft x-ray scattering~\cite{Ghiringhelli_Science2012,Blanco_PRL2013,Achkar_PRL2012}. As shown in Fig.~\ref{PhaseDiagram} the onset temperature of the Raman signal is always observed at a temperature significantly higher than the onset of the incommensurate signal seen with x-rays in the same samples~\cite{Ghiringhelli_Science2012,Blanco_PRL2013,Achkar_PRL2012}. This offset has to be related to the frequency of the probe. Recent hard x-ray experiments performed with high energy resolution~\cite{Blackburn_PRB2013, Letacon_NatPhys2013} have demonstrated that the x-ray intensity is dominated by a quasi-elastic ``central peak'' with energy width lower than $\sim$ 0.1 meV, corresponding to fairly slow ($\sim$ 13 ps), quasi-static, fluctuations of the charge density in the system, which appears in the 150-170 K range for doping levels close to $p = 0.11$. 
Phonons on the other hand, can be seen as probes of the charge dynamics at higher frequency, \textit{i. e.} on a much faster time scale ($\sim$ 0.1 ps). The appearance of the new modes around 220 K in YBa$_{2}$Cu$_{3}$O$_{6.6}$ is hence a natural consequence of the survival of fast CDW fluctuations up to this temperature.

In Fig.~\ref{PhaseDiagram} we have plotted the onset temperature of the new phonons as a function of doping, along with other phase transitions or crossover lines reported in the literature. The onset temperatures reported for the pseudogap differ substantially, possibly caused by the variety of criteria used to identify this phenomenon in different experimental techniques. Nonetheless, these temperatures can be roughly grouped into two categories~\cite{Chang_NaturePhysics2012,Hosur_PRB2013,Shekhter_Nature2013,Alloul_arxiv}. The first one (labeled $T^*_1$ in Fig.~\ref{PhaseDiagram})
corresponds to the reported doping dependence of the anomalous polar Kerr effect~\cite{Kapitulnik_NJP2009} in YBa$_{2}$Cu$_{3}$O$_{6+x}$ single crystals, and also coincides approximately with the sign change of the Hall coefficient~\cite{Leboeuf_PRB2011} and the decrease of the drop of the dynamical magnetic susceptibility seen in NMR relaxation measurements~\cite{Warren_PRL89, Berthier_PhysicaC1997, Baek_PRB2012}. The other line labeled $T^*_2$, at higher temperature refers, among others, to the drop of the static spin susceptibility seen in NMR Knight shift measurements~\cite{Alloul_PRL1989}, the onset of the $Q=0$ polarized neutron diffraction signal~\cite{Fauque_PRL2006, Mook_PRB2008, Baledent_PRB2011}, the signature of a broken rotational symmetry in the Nernst effect~\cite{Daou_Nature2010}, the decrease of spectral weight in the c-axis optical conductivity~\cite{Homes_PRL93, Dubroka_PRL2011}, and the more recently observed anomalies in resonant ultrasound spectroscopy~\cite{Shekhter_Nature2013}. These two lines differ by $\sim$ 100 K around 10$\%$ doping. The onset temperatures of the Raman phonons and of the quasi-static CDW fluctuations generally fall in between $T^*_1$ and $T^*_2$, but neither of them matches the doping dependence of the these lines. In any event, the CDW fluctuations appear below $T^*_1$, indicating that the CDW might be an instability within the pseudogapped state. Further work is required to elucidate the relationship between the CDW and the pseudogap determined by other spectroscopic probes.

At optimal doping,  no evidence for the CDW-related Raman phonons is discernible in our data down to 10 K.
Looking carefully at the data obtained on the most underdoped YBa$_{2}$Cu$_{3}$O$_{6.45}$ sample, it appears that at least the 205 and 560 cm$^{-1}$ features are visible. However, the ratio of their intensity to the one of the 340 cm$^{-1}$ phonon is an order of magnitude weaker than in the YBa$_{2}$Cu$_{3}$O$_{6.55}$ and YBa$_{2}$Cu$_{3}$O$_{6.6}$ compounds. At the same time, quasi-static incommensurate magnetism is observed using INS~\cite{Hinkov_Science2008,Haug_PRL2009} and NMR~\cite{Wu_PRB2013,Baek_PRB2012} at low temperature in the same compound. This confirms that the two types of modulations (spin and charge) compete with each other in the 123 family~\cite{Blanco_PRL2013}. Clearly, the charge modulation replaces the spin modulation precisely in the doping region where, at high magnetic fields, the metal-insulator transition dissapear and quantum oscillations have been observed ~\cite{Sebastian_PNAS2010}.


\section{Conclusions}

In conclusion, using Raman scattering we have observed clear signatures of the CDW state recently discovered by x-ray scattering  in underdoped single crystals of YBa$_2$Cu$_3$O$_{6+x}$. This observation allowed us to study the doping dependence of this new phenomenon in detail. The results highlight the subtle interplay between the various phases competing with superconductivity in the underdoped cuprates. Owing to the versatility of Raman scattering, the identification of Raman signatures of CDW formation will facilitate further investigations of the CDW under extreme conditions such as high magnetic fields and high pressure.

\section*{Acknowledgement}

We acknowledge fruitful discussions with C.~Bernhard, L. Boeri,  M. Cardona, M. Calandra, T. P. Devereaux, D.~Efremov, G.~Khaliullin, Y.~Li, D. Manske, I. Mazin, A.~Schnyder, and R.~Zeyher. M.B. thanks V.~Hinkov, D.~Haug, R.~Merkle, M.~Raichle and B.~Baum for their help in sample preparation, A.~Schulz for technical help, and the International Max Planck Research School for Advanced Materials for financial support.


\end{document}